\documentclass[aps,prl,showpacs,superscriptaddress,groupedaddress]{revtex4}  
\usepackage{graphicx}  
\usepackage{dcolumn}   
\usepackage{bm}        
\usepackage{amssymb}   
\usepackage{amsmath}
\usepackage{graphicx}
\usepackage{pdfpages}
 \pdfoutput=1

\usepackage{array}

\begin{document}

\title{Solubility of nitrogen in methane, ethane, and mixtures of methane and ethane at Titan-like conditions: A molecular dynamics study}

\author{Pradeep Kumar$^{1,2}$}
\email{pradeepk@uark.edu}
\author{Vincent F. Chevrier$^2$}
\affiliation{$^1$Department of Physics, University of Arkansas, Fayetteville, AR, 72701. \\ $^2$Arkansas Center for Space and Planetary Sciences, University of Arkansas, Fayetteville, AR, 72701.}

\date{\today}

\begin{abstract}
We have studied the temperature dependence of the solubility of nitrogen in methane, ethane, and mixtures of methane and ethane using vapor-liquid equilibrium simulations of binary and ternary mixtures of nitrogen, methane and ethane for a range of temperatures between $90$K and $110$K at a pressure of $1.5$~atm, thermodynamic conditions that may exist on the Saturn's giant moon, Titan. We find that --(i) the solubility of nitrogen in both methane and ethane decreases with increasing temperature;  (ii)  the solubility of nitrogen in methane is much larger compared to that in ethane at low temperatures, (iii) solubility of nitrogen in a ternary mixture of methane, ethane, and nitrogen increases upon increasing mole-fraction of methane. Our results are in quantitative agreement with the recent experimental measurement of the solubility of nitrogen in methane, ethane, and a mixture of methane and ethane. Furthermore, we find a strong temperature-dependent surface adsorption of nitrogen at the nitrogen-hydrocarbon interface, previously unknown. We have also investigated surface tension of the gas-liquid interface and find that it decreases upon decreasing temperature. Moreover, we find that the interfacial layer of adsorbed nitrogen and ethane show a preferential orientational ordering at the interface.
\end{abstract}
\maketitle
\noindent \section{Introduction}
Besides earth, Saturn's giant moon Titan is the only other planetary body in our solar system that has stable and accessible liquid on its surface, an active hydrologic cycle similar to Earth, and a dense atmosphere with a pressure of about $1.5$ times that of Earth~\cite{Lunine:2008aa,Lorenz:2010aa,Hayes:2018aa}. Early interest in a methane cycle on Titan~\cite{Flasar:1983aa,Toon:1988aa} was motivated by Voyager 1's discovery of a thick, nitrogen-based atmosphere with a significant methane abundance.  Ethane clouds in the Stratosphe were later identified by the Cassini  at high northern latitudes~\cite{Lunine:1983aa,Brown:2008aa}. Cassini mission to Saturn has further revealed many features of this moon including the existence of large liquid hydrocarbon lakes and seas composed primarily of methane and ethane in different ratios~\cite{Cordier2009}.  With the low surface-temperature of about $90$K and the pressure about $1.5$~atm, both methane and ethane condense out of the atmosphere and exist in the stable liquid phases on the surface in dynamic equilibrium with its atmosphere~\cite{Lorenz:2010aa}.

Cassini's Radio Detection and Ranging (RADAR) observations further suggested transient bright features on the second largest hydrocarbon sea, Ligeia Mare, known as "{\it Magic Islands}"
~\cite{Hofgartner:2014aa,HOFGARTNER2016338}. Hofgartner et. al. have argued that these features could not be image artifacts or permanent geophysical structures but are consistent with ephemeral phenomena such as suspended solids, bubbles, waves and tides on Titan\cite{HOFGARTNER2016338}. Using a numerical model~\cite{Cordier:2017aa} and existing solubility data of nitrogen in methane and ethane~\cite{Llave:1985aa,Llave:1987aa}, Cordier et. al. have hypothesized that these transient bright features may correspond to dissolution of nitrogen, giving rise to large nitrogen bubbles detected as bright objects by Cassini's RADAR.  Motivated by this,  laboratory measurements of solubility of nitrogen in liquid methane, ethane, and mixtures of methane and ethane, the primary components of seas and lakes on Titan, have been carried out~\cite{MALASKA201794,Cheverier:2018aa}. They find that the nitrogen's solubility in methane is much larger as compared to ethane. Furthermore, the solubility of nitrogen in both methane and ethane decreases upon decreasing temperature. Based on these results, various scenarios regarding the exsolution of nitrogen from the lakes has been proposed including exsolution of nitrogen with increasing temperature and exsolution of nitrogen when a methane-rich region meets an ethane-rich region. While thermodynamic models have been developed and experiments have been performed, a microscopic picture of the dissolution of nitrogen is still lacking.  Standard methods for calculating solubility in the limit of infinite dilution is Widom insertion method ~\cite{widom1966random}. Since the solubility of nitrogen is very high in both methane and ethane, we have performed molecular dynamics vapor-liquid equilibrium (VLE) simulations~\cite{buldyrev2007water,Minkara:2018aa,panagiotopoulos2002direct,panagiotopoulos1987direct,potoff2001vapor,morrow2019vapor} of binary mixtures of methane and nitrogen, ethane and nitrogen as well as ternary mixtures of methane, ethane, and nitrogen to investigate the solubility of nitrogen in methane, ethane, and mixtures of methane and ethane at Titan-like conditions. The paper is organized as follows: In the method section, we describe our simulation approaches for all the thermodynamic conditions studied here. In the Results section, we present the results for both the binary and ternary systems. Finally, we conclude our studies with summary and discussion.

\noindent\section{Method}
\medskip

\begin{figure}
\begin{center}
\includegraphics[width=14cm]{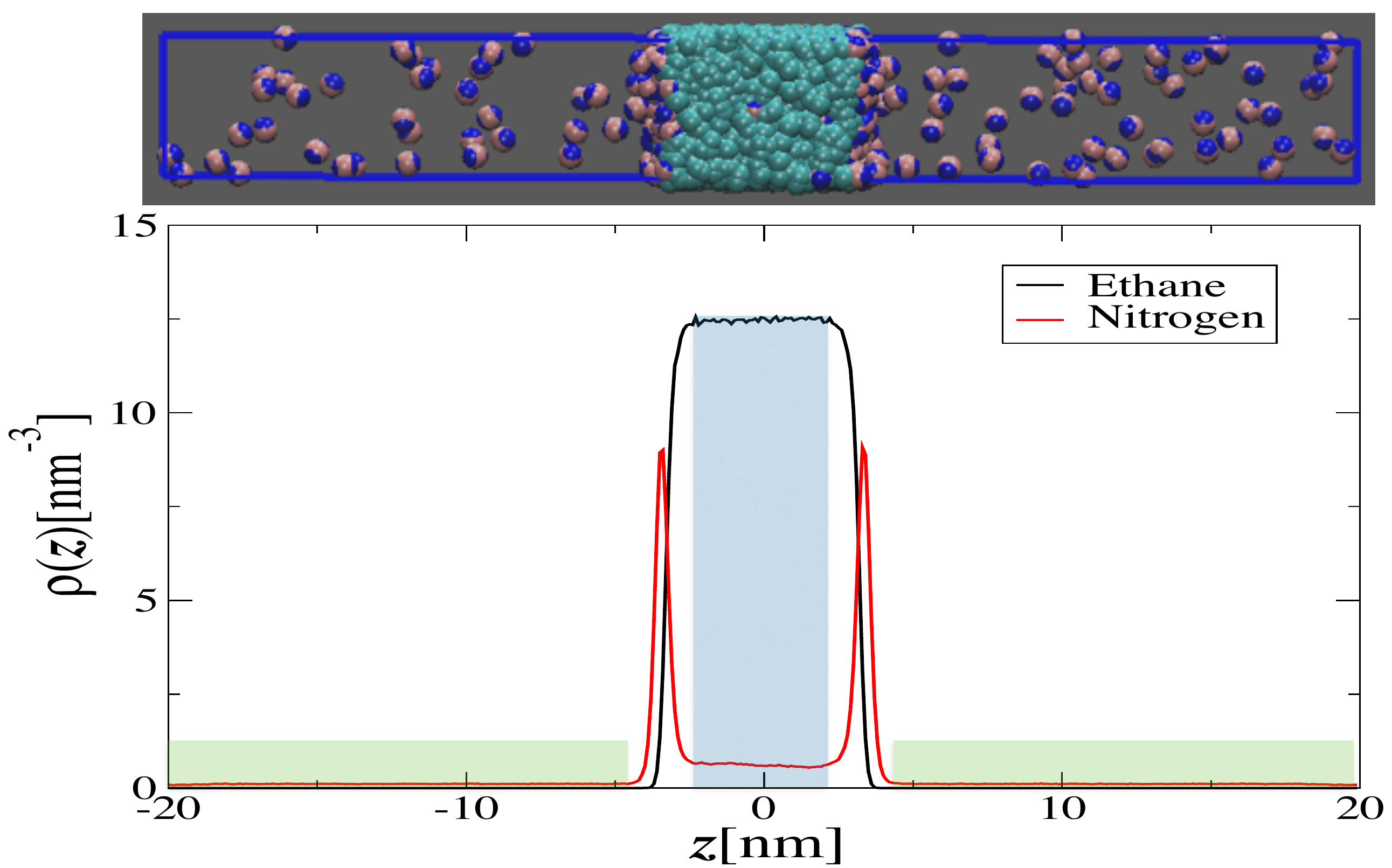}
\end{center}
\caption{A snapshot of the nitrogen-ethane binary mixture at $T=90$K (Top). Average number density profile, $\rho(z)$,  for ethane and nitrogen along the $z$-direction (Bottom). The blue shaded represents the liquid phase region and the green shaded region represents the vapor phase.}
\label{fig:fig1}
\end{figure}
We have carried out vapor-liquid  equilibrium (VLE) simulations of (i) binary mixtures of nitrogen and methane, (ii) nitrogen and ethane for a range of temperatures between $90$K and $110$K, and (iii) ternary mixtures of nitrogen, methane, and ethane containing different fractions of methane and ethane at temperature $T=90$~K.  All the simulations were performed in Gromacs4.6.5~\cite{BERENDSEN199543,Lindahl:2001aa,Van-Der-Spoel:2005aa}. The trappe-UA force field~\cite{keasler2012transferable,wick2005transferable,wick2000transferable,martin1999novel} was used to model methane, and ethane was modeled using an improved parameterization of ethane, trappe-UA2~\cite{shah2017transferable}. Trappe-small parameterization was used to model nitrogen. Lorent-Bertloth~\cite{frenkel2001understanding} rule was used to model the cross interactions of nitrogen with methane and ethane. The short-range van der Wall interactions were treated with a  cut-off of $1.5$~nm and particle-mesh-Ewald(PME)~\cite{frenkel2001understanding} was used for the long range interactions.  Since the system is not homogeneous, dispersion corrections to the energy and pressure were not applied. First, the liquid phase simulations of methane or ethane was performed at pressure corresponding temperatures and at pressure, $P=1.5$~atm. Separately a gas-phase simulation was performed with nitrogen at the same pressure and temperature and subsequently the simulation boxes of two nitrogen and methane was put together to form a simulation box consisting of nitrogen-ethane system as shown in Figure 1. The dimension of the final simulation box was $Lx=Ly=5.0$nm$< L_z$. In such a box the interface is stable and forms along the smallest surface area in the $xy$-plane, perpendicular to the long-axis. The number of molecules of methane and ethane was fixed to $3000$ and $2000$ for all the binary mixture simulations (nitrogen-methane, nitrogen-ethane) and the number for nitrogen molecules varied for different temperatures depending on the solubility and the gas phase density. If the same number of molecules of nitrogen are used for the temperatures where the solubility is small, the box size would be enormously large and the computational load of the Ewald summation would be very high.  The equations of motion are integrated with a time step of $2$~fs and velocity rescaling is  used to attain constant temperature and anisotropic Berendsen barostat for constant pressure, $P_{zz}=1.5$~atm in the $z$-direction.  After the equilibration for $60$~ns, we ran the simulations for additional $80$~ns for each state point  and the equilibrium averages are calculated from these trajectories.

\section{Results}

\noindent {\bf \large Solubility of nitrogen in methane and ethane}
\medskip

\begin{figure}
\begin{center}
\includegraphics[width=16cm]{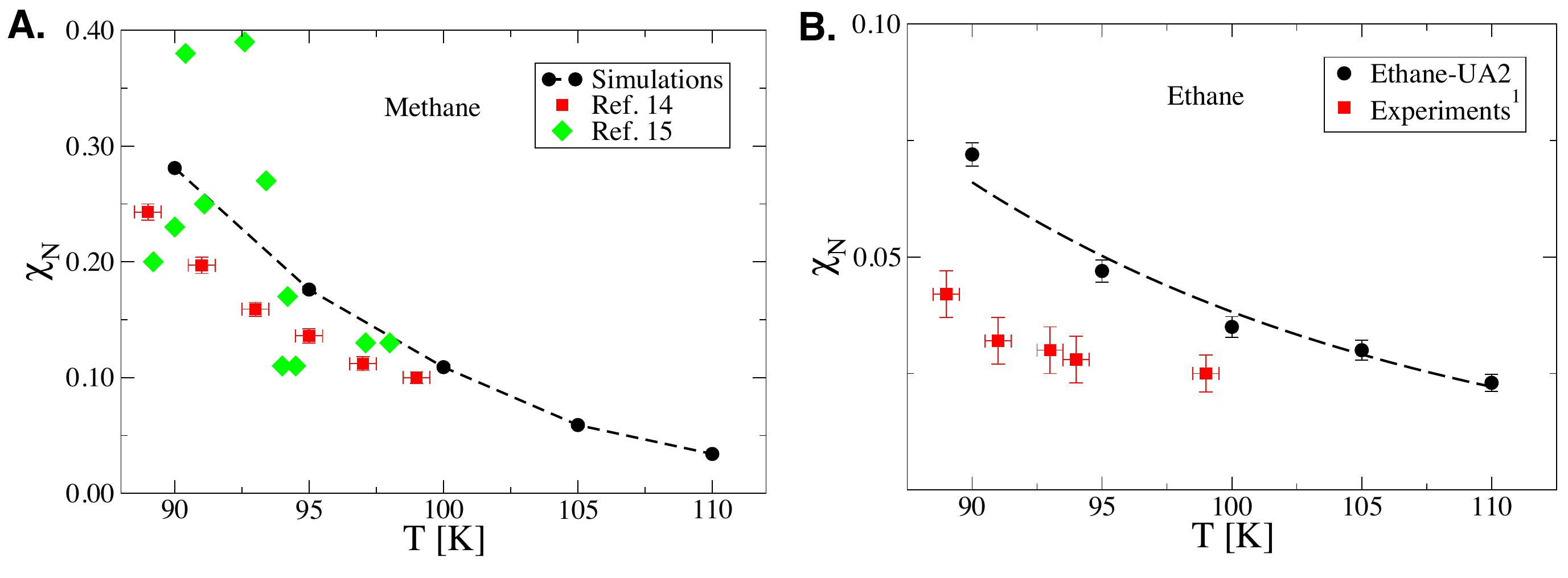}
\end{center}
\caption{(A) Mole-fraction, $\chi_N$, of the nitrogen in methane as a function of temperature. To compare with experiments, we also show the data from Refs.~\cite{MALASKA201794,farnsworth2017experimental}. Simulation results are in reasonable agreement with the experimental data with a slight overestimation of solubility at low temperatures. (B) Mole-fraction, $\chi_N$, of the nitrogen in ethane as a function of temperature.  For a comparison, we also plot the data from Ref.~\cite{MALASKA201794}.  Nitrogen in ethane exhibits a decrease of solubility with temperature and are in quantitative agreement with experimental values in Ref.~\cite{MALASKA201794}. The error in the solubility is estimated from the errors in the mole fractions in the liquid phase.} 
\label{fig:fig2}
\end{figure}

To compute the solubility, we measure the mole-fraction of nitrogen in the liquid phase of methane/ethane in the nitrogen-methane and nitrogen-ethane binary mixtures at equilibrium. To avoid the interface, we define the liquid phase (or the gas phase) as the region where the $z$-derivative of the density of methane/ethane and nitrogen is zero (ses Fig.~\ref{fig:fig1}). We count the number of molecules of nitrogen, $\mathcal{N}_N^{\ell}$, in the liquid phase of methane/ethane and similarly count the number of molecules of methane/ethane, $\mathcal{N}_M^{\ell}$ or $\mathcal{N}_E^{\ell}$, in the liquid-phase for the corresponding binary mixture simulations. The solubility as measured by the mole-fraction, $\chi_N$, of nitrogen in the liquid phase is defined as
\begin{align}
\chi_{N} &= \frac{\mathcal{N}_N^{\ell}}{\left(\mathcal{N}_N^{\ell}+N_M^{\ell}\right)} \text{ For nitrogen-methane binary system} \\
	     &= \frac{\mathcal{N}_N^{\ell}}{\left(\mathcal{N}_N^{\ell}+N_E^{\ell}\right)}  \text{ For nitrogen-ethane binary system}
\end{align}
In Fig.~\ref{fig:fig2}(A), we show the solubility of nitrogen in methane for temperatures $T=90,95,100,105$, and $110$K respectively. Consistent with the experimental results, we find that the solubility decreases upon increasing temperature. To compare the simulation results with experiments, we also show the data from two different experiments~\cite{MALASKA201794,farnsworth2017experimental,Cheverier:2018aa}. The solubility values calculated in our simulations are very close to experimental values with small deviations at lower temperatures. The experimentally measured value of solubility at $T=90$K is $0.235\pm0.005$~\cite{MALASKA201794} as compared to the value $0.278+0.002$ in our simulations. In Fig.~\ref{fig:fig2}(B), we show the Mole-fraction, $\chi_N$, of the nitrogen in ethane as a function of temperature.  For a comparison, we also plot the data from Ref.~\cite{MALASKA201794}.  Nitrogen in ethane exhibits a decrease of solubility with temperature and are in quantitative agreement with experimental values. The solubility values of nitrogen for the nitrogen-methane and nitrogen-ethane systems are also listed in Table~I.

\bigskip

\newcommand{\PreserveBackslash}[1]{\let\temp=\\#1\let\\=\temp}
\newcolumntype{C}[1]{>{\PreserveBackslash\centering}p{#1}}
\begin{center}
\begin{table}
\begin{tabular}{c|C{2cm} | C{2cm} | C{2cm} |C{2cm}|C{2cm}|C{2cm}|C{2cm}}
\hline
& $\chi_M $ &$\Delta \chi_M$& $\chi_E $ &$\Delta \chi_E$ & $\chi_N$ & $\Delta \chi_N$ &$T$ (K) \\
 \hline
 {\bf M}&1.00 	&& 0.00&	&0.278 & 	0.002& 	90\\
 {\bf E}&1.00		&& 0.00& 	& 0.176&	0.003 	&	95\\
 {\bf T}&1.00		&& 0.00&	&0.109 &	0.002	&	100\\
 {\bf H}&1.00		&& 0.00& 	& 0.059 &	 0.002  	&	105\\  
 {\bf A}&1.00		&& 0.00&	& 0.034&	0.001 	& 	110\\
{\bf N}&&&&&&&\\
{\bf E}&&&&&&&\\
\hline
{\bf E}& 0.00		&& 1.00&	&0.072&	0.003	& 	90\\
{\bf T}& 0.00		&& 1.00&	&0.047&	0.002	& 	95\\
{\bf H}& 0.00		&& 1.00&	&0.035&	0.002	& 	100\\
{\bf A}& 0.00		&& 1.00&	&0.035&	0.002	& 	105\\
{\bf N}& 0.00		&& 1.00&	&0.031&	0.002	& 	110\\
{\bf E}&&&&&&&\\ 
\hline
	 &			&		&		&			&				&				&\\
{\bf M}&  0.067		&0.002	&0.855 	&	0.002	&0.077			&	0.002		& 	90\\
{\bf I}& 0.139		&0.002	& 0.777	&	0.002	&0.083			&	0.003		& 	90\\
{\bf X}& 0.290		&0.002	&0.594	&	0.002	&0.115			&	0.002		& 	90\\
{\bf T}& 0.371		&0.002	&0.490	&	0.002	&0.138			&	0.002		& 	90\\
{\bf U}& 0.448		&0.002	& 0.391	&	0.002	&0.160			&	0.002		& 	90\\
{\bf R}& 0.528		&0.002     & 0.282	&	0.002	&0.189			&	0.002		& 	90\\
{\bf E}&0.595 		&0.002	& 0.182	&	0.002	&0.222			&	0.002		& 	90\\
	 &0.659 		&0.002	& 0.086	&	0.002	&0.254			&	0.002		& 	90\\

\hline
\end{tabular}
\label{tab:table1}
\caption{Mole fraction of methane, $\chi_M$, ethane, $\chi_E$, and nitrogen, $\chi_N$ for nitrogen-methane, nitrogen-ethane, and nitrogen-methane-ethane. The mole fractions $\chi_M$ and $\chi_E$ for the ternary mixture are the liquid-phase mole fractions. We also list the estimated errors, $\Delta \chi_{X}$ for all the state points studied here.}
\end{table}
\end{center}

\bigskip

\noindent{\bf \large Surface Tension}
\medskip

\begin{figure}
\begin{center}
\includegraphics[width=16cm]{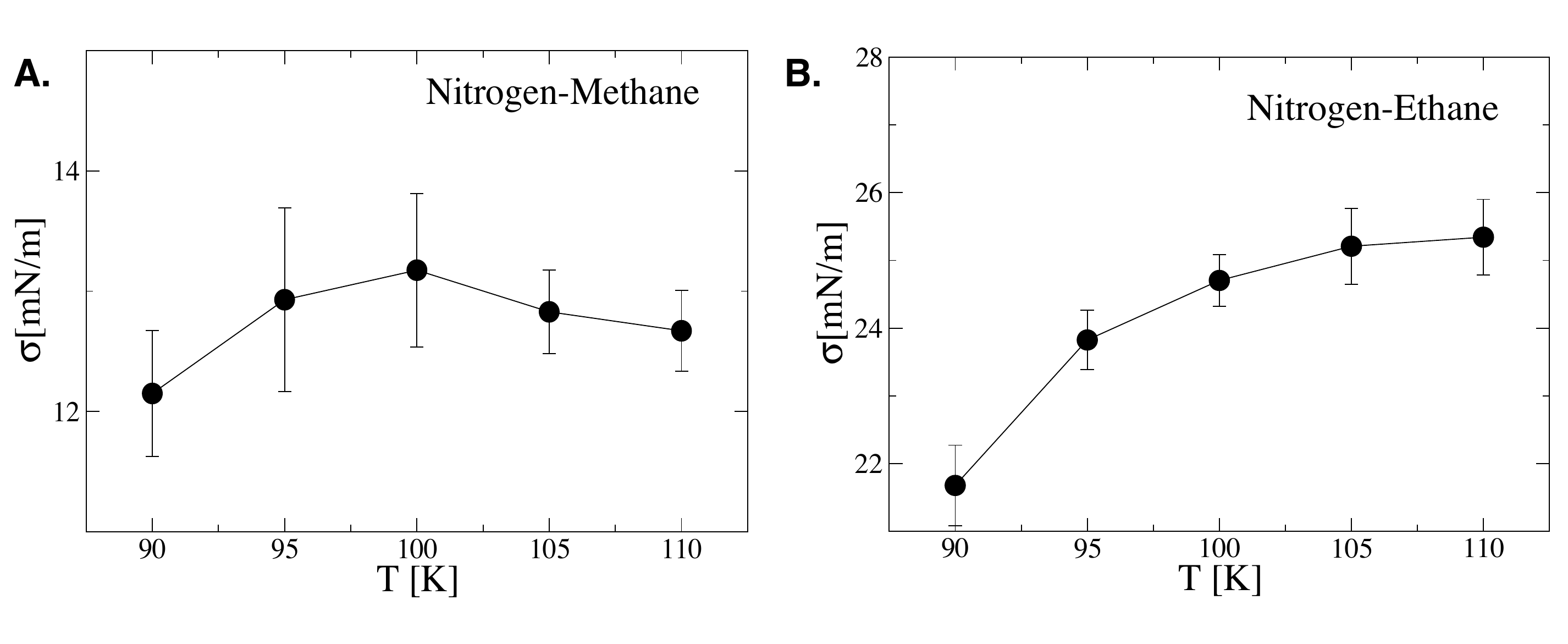}
\end{center}
\caption{Temperature dependence of the surface tension, $\sigma$, for (A) nitrogen-methane and (B) nitrogen-ethane systems. Surface tension for the nitrogen-methane system is smaller as compared to surface tension of the nitrogen-ethane system. Moreover, the surface tension for the nitrogen-ethane system decreases monotonically upon decreasing temperature, while the surface tension for the nitrogen-methane system exhibits a non-monotonic behavior with temperature.}
\label{fig:fig3}
\end{figure}

\begin{figure}
\begin{center}
\includegraphics[width=16cm]{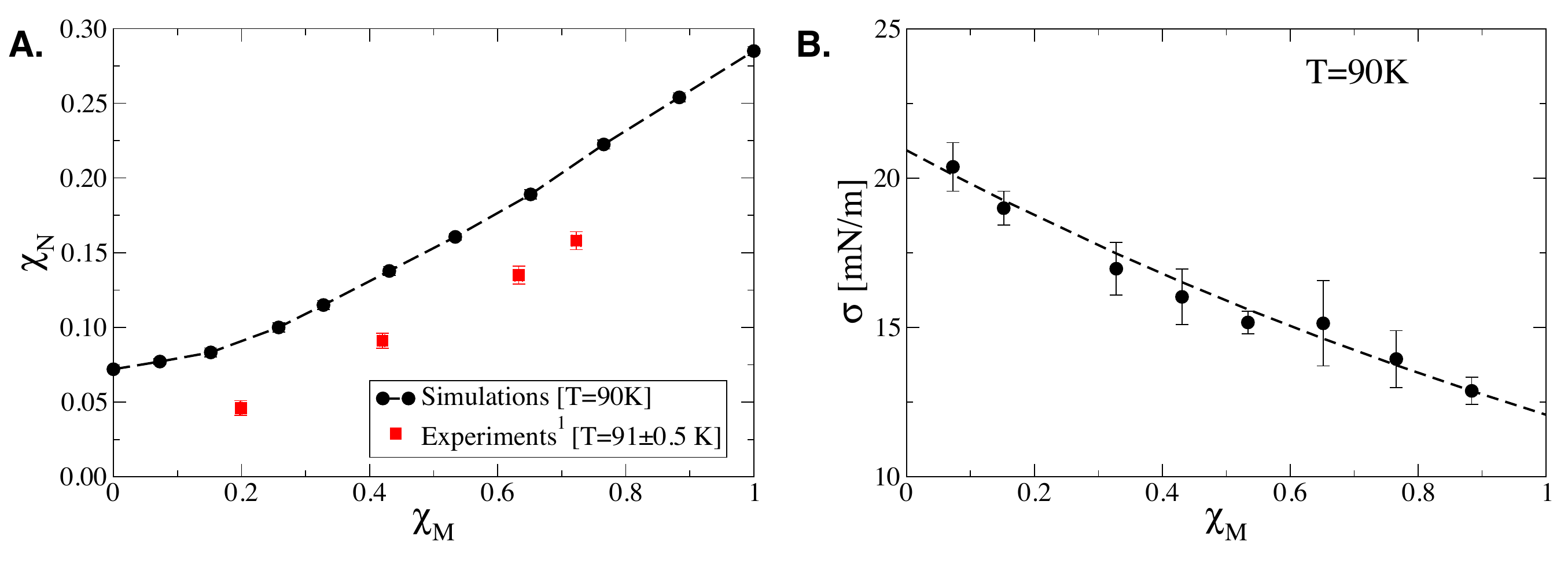}
\end{center}
\caption{(A) Solubility of nitrogen as a function of mole-fraction of methane in the liquid state for the nitrogen-methane-ethane system. For comparison, we also plot the data from Ref.~\cite{MALASKA201794} at $T=91\pm0.5K$.  We find that solubility of nitrogen increases with increasing mole-fraction of methane in the liquid-phase, similar to the experimental observations~\cite{MALASKA201794}.  (B) Surface tension as a function of mole fraction of methane in the liquid state in a mixture of methane and ethane. Solubility of nitrogen increases as the mole fraction of methane in the mixture increases while the surface tension decreases upon increasing mole fraction of methane. Our values of surface tension are agree well with the experimental values~\cite{baidakov2011capillary,baidakov2012surface,baidakov2016surface,blagoi1960surface}.}
\label{fig:fig4}
\end{figure}

To understand the temperature dependence of the solubility, we next studied the temperature dependence of the surface tension, which is readily available from the molecular dynamics simulations. The surface tension, $\sigma$, is defined as
\begin{equation}
\sigma = \frac{L_z}{2}\left[P_{zz} -0.5(P_{xx}+P_{yy})\right]
\end{equation}
where $L_z$ is the box-length in the $z$-direction and $P_{xx}$, $P_{yy}$, $P_{zz}$ are the diagonal components of the pressure tensor in the $x$, $y$, and $z$-directions, respectively. A factor of $2$ accounts for the presence of two interfaces in the simulation box. Figures ~\ref{fig:fig3} (A) and (B) show the temperature dependence of the surface tension, $\sigma$, for the nitrogen-methane and the nitrogen-ethane binary mixtures, respectively. We find that the surface tension of nitrogen-methane interface is about $8mN/m$ smaller than the surface tension of the nitrogen-ethane interface for all the temperatures investigated here. Furthermore, we find that $\sigma$ decreases with decreasing temperature for the nitrogen-ethane system. For the nitrogen-methane system,  surface tension shows a non-monotonic dependence on temperature owing to larger partial pressure of methane at $T\ge105$K. The partial pressure of ethane is negligible for all the temperatures studied here. Indeed, we don't see a single molecule of ethane in the vapor phase over the time scale of our simulations. Our results for surface tension values are in quantitative agreement with the experimental data~\cite{baidakov2011capillary,baidakov2012surface,baidakov2016surface,blagoi1960surface} for both nitrogen-methane and nitrogen-ethane systems.

\bigskip

\noindent {\bf \large Solubility of nitrogen in a mixture of methane and ethane}

\bigskip

We next studied the solubility of nitrogen in a ternary mixture of methane, ethane, and nitrogen containing different fractions of methane/ethane at $T=90$~K and $P=1.5$~atm.  In Fig.~\ref{fig:fig4}(A), we show the solubility of nitrogen for the nitrogen-methane-ethane system as a function of different mole-fraction of methane in the liquid-phase. Since the partial pressure of methane at this temperature and pressure is non-zero, the mole-fraction of methane in the liquid-phase is not the same as the fraction of methane in the simulation box. The methane mole-fraction, $\chi_M$, in the liquid phase is defined as
\begin{equation}
\chi_{M} = \frac{\mathcal{N}_M^{\ell}}{\mathcal{N}_M^{\ell}+\mathcal{N}_E^{\ell}}
\end{equation}
where, $\mathcal{N}_M^{\ell}$, $\mathcal{N}_E^{\ell}$, and $\mathcal{N}_N^{\ell}$ are the numbers of molecules of methane, ethane, and nitrogen in the liquid-phase, respectively. Similarly the solubility as measured by the mole-fraction, $\chi_N$, of nitrogen in the liquid-phase is defined as
\begin{equation}
\chi_{N} = \frac{\mathcal{N}_N^{\ell}}{\mathcal{N}_M^{\ell}+\mathcal{N}_E^{\ell}+\mathcal{N}_N^{\ell}}
\end{equation}
To compare our simulation results with experiments, we also show the solubility data from Ref.~\cite{MALASKA201794}. We find that solubility of nitrogen increases with increasing mole-fraction of methane in the liquid-phase, similar to the experimental observations~\cite{MALASKA201794}.  Simulation results are in good agreement with the experimental data with slight overestimation of the solubility for all the mole-fractions of methane. In Fig.~\ref{fig:fig4}(B), we show the behavior of surface tension as a function of mole-fraction of methane in the liquid-phase. We find that the surface tension decreases with increasing mole-fraction of methane. The solubility values of nitrogen for the nitrogen-methane-ethane system are summarized in Table~I.

\bigskip

\noindent {\bf \large Adsorption of nitrogen at the interface}
\bigskip

\begin{figure}
\begin{center}
\includegraphics[width=16cm]{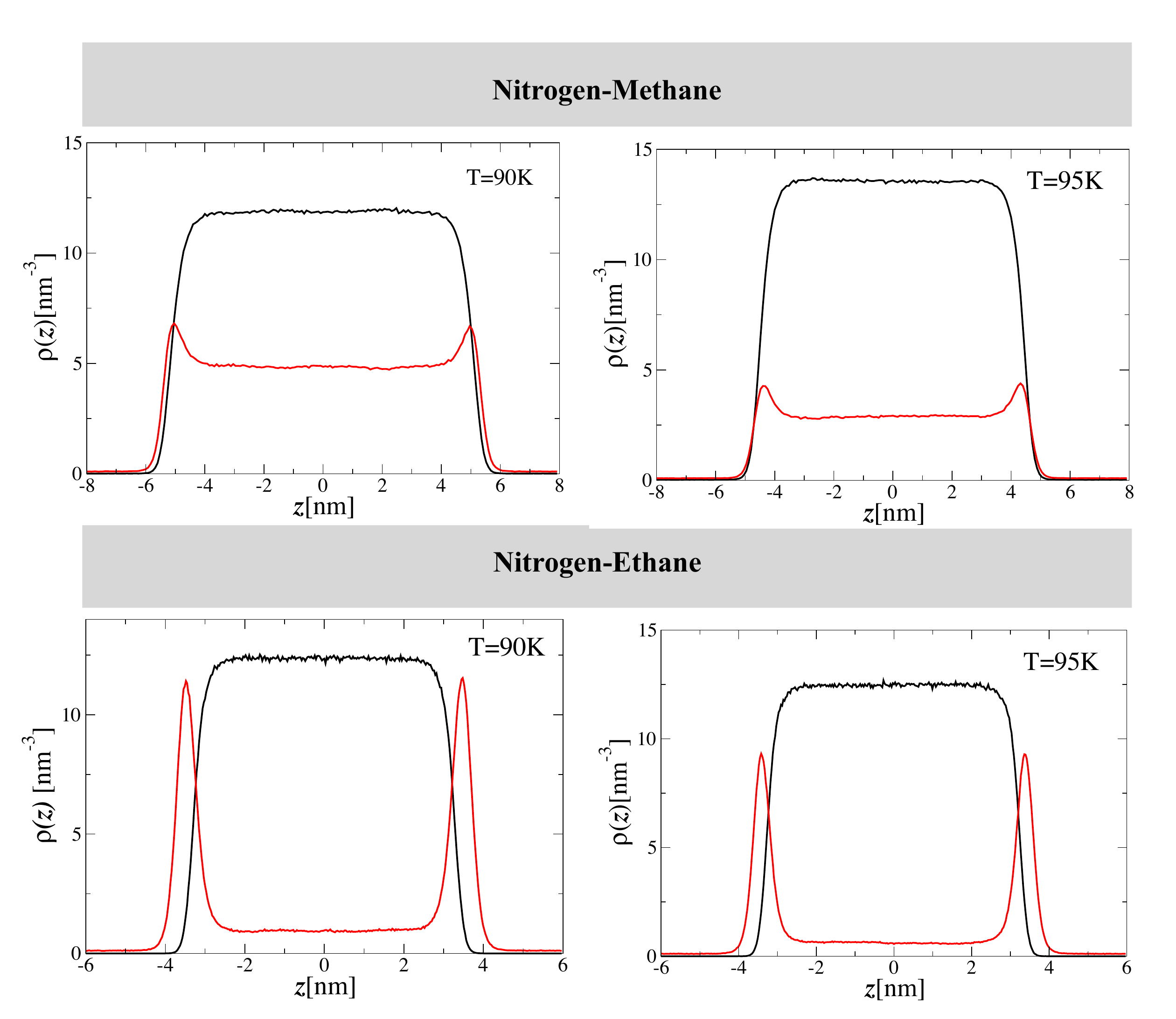}
\end{center}
\caption{Number density, $\rho(z)$, of nitrogen (solid red curve) and methane/ethane (solid black curve) for the nitrogen-methane and nitrogen-ethane system for two different temperatures $T=90$~K and $T=95$~K. A strong temperature-dependent surface adsorption of nitrogen is observed in both systems.}
\label{fig:fig5}
\end{figure}
\begin{figure}
\begin{center}
\includegraphics[width=16cm]{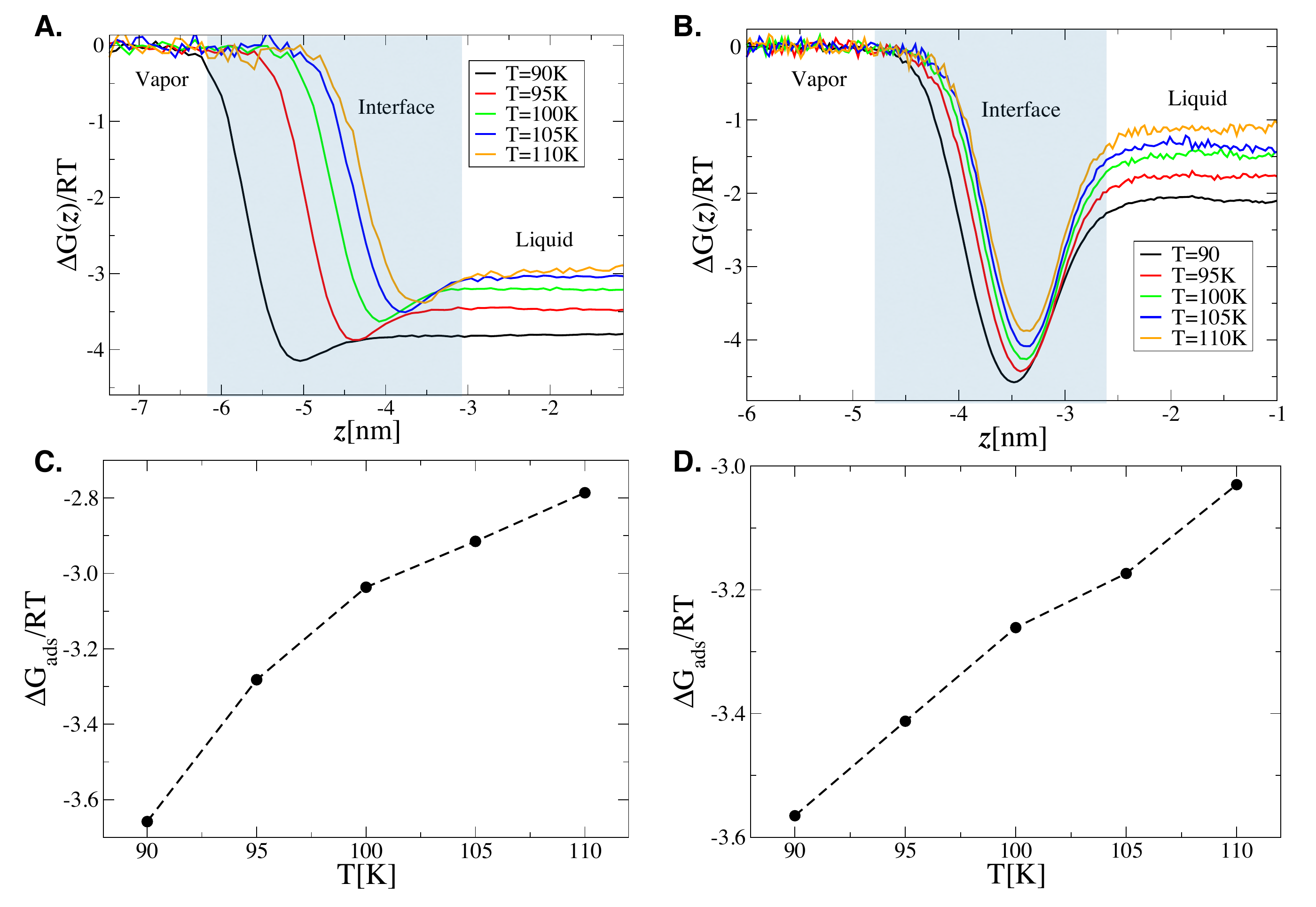}
\end{center}
\caption{Free energy profile, $\Delta G(z)$, at different temperatures for (A) nitrogen-methane and (B) nitrogen-ethane system. The data is only shown for the $z$-values close to the interface so that one can observe the vapor, the interface and the liquid regions. Free energy of adsorption, $\Delta G_{\rm ads}/RT$, as a function of temperature for (A) nitrogen-methane, and (D) nitrogen-ethane systems.}
\label{fig:fig6}
\end{figure}

In this section, we investigate the adsorption of nitrogen at the nitrogen-methane and the nitrogen-ethane interface. In Fig.~\ref{fig:fig5}, we show the density profile of nitrogen and methane/ethane as a function of temperature for both nitrogen-methane and nitrogen-ethane systems at two different temperatures $T=90$K and $T=95$K . We find that adsorption of nitrogen at the interface between vapor and liquid phase is very high and increases upon decreasing temperature. Furthermore, the degree of adsorption of nitrogen at the nitrogen-ethane interface is much higher as compared to the nitrogen-methane interface whereas the number density of nitrogen reaches approximately the number density of liquid ethane. Many other studies of liquid-gas interfaces have also found a strong adsorption of gases at the gas-liquid interface~\cite{masterton1963surface,Minkara:2018aa}. 

The partition coefficient, $K$, of two phases is defined as the ratio of the number density of the phases. Here we can define a $z$-dependent partition coefficient, $K(z)$ 
\begin{equation}
K(z) = \frac{\rho(z)}{\rho^v}
\end{equation}
where $\rho(z)$ is the density profile along the $z$-direction and $\rho^v$ is the density of the nitrogen in the vapor phase. Consequently, one can define excess free energy, $\Delta G(z)$, over the free energy of the vapor-phase as
\begin{equation}
\Delta G(z) = -RT\log K(z)
\end{equation}
where $R$ is the universal gas constant and $T$ is the temperature. In Figs.~\ref{fig:fig6}(A)(B), we show $\Delta G(z)$ for different temperature for nitrogen-methane and nitrogen-ethane systems We find that $\Delta G(z)$ decreases upon entering the adsorbate region and increases slightly and levels off in the liquid region, suggesting the free change in adsorbate region is larger compared to that in the liquid-region and hence adsorption at the interface. Moreover, we find that the value of $\Delta G(z)$ decreases upon decreasing temperature and hence stronger adsorption at lower temperatures. To find the free energy associated with the exchange with the adsorbate region, $\Delta G_{\rm ads}$, one must define the effective density, $\bar{\rho}_I$, of the adsorbate region. We define the adsorbate region as the region between the vapor and liquid phases in which the derivative of nitrogen-density is non-zero. This definition of characterizing interface, gas, and liquid regions are adopted from Ref.~\cite{buldyrev2007water} . $\Delta G_{ads}$ is defined as
\begin{equation}
\Delta G_{ads} = -RT \log{\frac{\bar{\rho}_I}{\rho^{v}}}
\end{equation}
 In Figs.~\ref{fig:fig6}(C)(D), we show $\Delta G_{ads}$  as a function of temperature for nitrogen-methane and nitrogen-ethane systems, respectively. We find that $\Delta G_{\rm ads}$ values for both nitrogen-methane and nitrogen-ethane are similar and decrease upon decreasing temperature, and consequently increasing propensity of adsorption of nitrogen at the interface. Note that the value of $\Delta G_{\rm ads}$ is smaller than the free energy difference between the gas and the liquid phase. We have summarized the values of $\Delta G_{\rm ads}$ at different temperatures for both systems in Table~II.
\begin{center}
\begin{table}
\begin{tabular}{|c | c | c|}
\hline
 T (K) &  $\Delta G_{\rm ads}/RT$ & $\Delta G_{\rm ads}/RT$\\
  &  Nitrogen-Methane & Nitrogen-Ethane\\
 \hline
 90 & -3.59  & -3.56\\
 95 & -3.21  & --3.41\\
100 & -3.00 & -3.26\\
105 &  -2.90& -3.17\\
110 & -2.77 & -3.02\\
\hline 
\end{tabular}
\label{tab:table2}
\caption{$\Delta G_{\rm ads}$ for nitrogen-methane and nitrogen-ethane systems, respectively.}
\end{table}
\end{center}

\bigskip

\noindent{\bf \large Orientational ordering of the ethane and nitrogen at the interface}
\medskip

\begin{figure}
\begin{center}
\includegraphics[width=14cm]{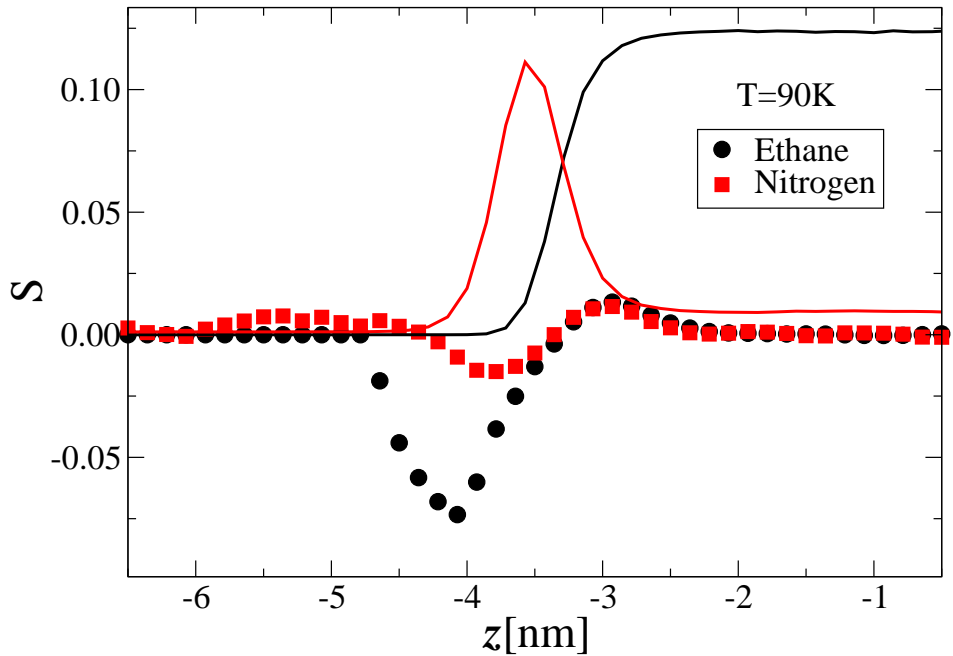}
\end{center}
\caption{Orientational order parameter, $S$, for ethane (black circles) and nitrogen (red squares) for nitrogen-ethane mixture at $T=90$K.  Also shown are the rescaled density profiles of nitrogen and ethane along the $z$-axis in red and black solid curves, respectively. For clarity we only show the region near the interface. Both ethane and nitrogen lying in the adsorbate region tend to have a preferential orientation minimum and a maximum in $S$ in the adsorbate region.}
\label{fig:fig7}
\end{figure}

We next investigate if the nitrogens and ethanes lying in adsorbate region exhibit any preferential orientational ordering. Recent studied of hexane-water system suggest that hexane tend to exhibit a preferred orientation in the adsorbate region~\cite{Minkara:2018aa}. To measure the degree of orientational ordering of of nitrogen and ethane at the interface, we use an orientational order parameter, S, as~\cite{Minkara:2018aa}
\begin{equation}
S = \frac{1}{2}\left < 3cos^2\theta -1\right>
\end{equation}
where $\theta$ is the angle formed by the nitrogen-nitrogen bond (nitrogen molecule) or carbon-carbon bond (ethane molecule) with the normal to the interface. This order parameter is closely related to the NMR order parameter~\cite{woessner1962nuclear,lemaster1999nmr,brueschweiler1994nmr}. If the molecules are oriented parallel to the interface then the value of $S$ reaches its maximum value $1$, and if they are oriented perpendicular to the interface, the value of $S$ reaches its minimum value $-0.5$. The value of $S$ is zero if they are oriented randomly or are oriented at the magic angle, $57.3^{\circ}$. In Fig.~\ref{fig:fig7}, we show the value of $S$ as a function of $z$ for nitrogen-ethane system at $T=90$K. We find that $S$ for nitrogen is zero both in the liquid and the vapor-phase suggesting random orientation of nitrogen in these phases. However, nitrogens are weakly aligned along the normal to the interface close to the gas-phase whereas $S$ has weak negative minimum and weakly aligned along interface in the region close to the liquid-phase. The value of $S$ for the ethane shows a similar behavior but with larger preference of alignment normal to the surface in the region close to the gas-phase. The value of $S$ for ethane in the liquid-phase is zero suggesting a random orientation of the molecules.
\section*{Summary and Discussion}
We have studied the temperature dependence of the solubility of nitrogen in methane, ethane, and mixtures of methane and ethane by performing extensive vapor-liquid equilibrium simulations of binary and ternary mixtures of nitrogen, methane and ethane for a range of temperatures between $90$K and $110$K at a pressure of $1.5$~atm, thermodynamic conditions that may exist on the Saturn's giant moon, Titan. We find that the solubility of nitrogen in both methane and ethane decreases with increasing temperature. Moreover,  solubility of nitrogen in methane is much larger compared to that in ethane at lower temperatures. Furthermore, we also investigated the solubility of nitrogen in a ternary mixture of methane, ethane, and nitrogen and we find that the solubility increases upon increasing mole-fraction of methane in the liquid phase. Our results are in quantitative agreement with the recent experimental measurements of the solubility of nitrogen in methane, ethane, and a mixture of methane and ethane. We have also calculated the surface tension for both binary and ternary systems and find the surface tensions measured in our simulations are in agreement with the existing data on the surface tensions of nitrogen-methane and as well nitrogen-ethane systems. We find a strong temperature-dependent surface adsorption of nitrogen at the nitrogen-hydrocarbon interface, previously unknown. The interfacial layer of adsorbed nitrogen and and the interfacial ethane show a preferential orientational ordering at the interface. Such adsorption of nitrogen might affect other processes like dissolution of atmospheric compounds in Titan lakes.

\section*{Author Contributions} PK designed, performed the research, and analyzed the data. PK and VFC wrote and revised the manuscript.

\section*{Acknowledgment}
Authors would like to thank University of Arkansas High Performance Computing Center for providing computational time. V. F. Chevrier acknowledges funding from NASA Cassini Data Analysis Program grant no. NNX15AL48G.

\bibliographystyle{unsrtnat}
\bibliography{thebibrev}

\end{document}